\documentclass[aps,prl,8pt,onecolumn,showpacs,amsmath,amssymb,floatfix,footinbib,superscriptaddress]{revtex4-1}
\usepackage{amsmath,natbib,graphicx,graphics,subfigure,amssymb,mathrsfs,CJK,color}
\usepackage{multirow,fancyhdr,color,bm,tabularx,psfrag,geometry,dcolumn,datetime}
\usepackage[colorlinks=true,bookmarksopen,bookmarksnumbered,citecolor=blue, linkcolor=blue, urlcolor=blue]{hyperref}
\def\footnoterule{\kern -1mm \hrule width 6.6cm \kern 2.2mm}%
\usepackage{tikz,xcolor,hyperref}
\definecolor{lime}{HTML}{A6CE39}
\DeclareRobustCommand{\orcidicon}{%
    \begin{tikzpicture}
    \draw[lime, fill=lime] (0,0)
    circle [radius=0.16]
    node[white] {{\fontfamily{qag}\selectfont \tiny ID}};\draw[white, fill=white] (-0.0625,0.095)
    circle [radius=0.007];
    \end{tikzpicture}
    \hspace{-2mm}}
\foreach \x in {A, ..., Z}
{\expandafter\xdef\csname orcid\x\endcsname{\noexpand\href{https://orcid.org/\csname orcidauthor\x\endcsname}{\noexpand\orcidicon}}}

\begin{document}

\title{Quantum entropy evolution in the photovoltaic process of a quantum dot photocell}
\author{Lin-Jie Chen }
\affiliation{Center for Quantum Materials and Computational Condensed Matter Physics, Faculty of Science, Kunming University of Science and Technology, Kunming, 650500, PR China}
\affiliation{Department of Physics, Faculty of Science, Kunming University of Science and Technology, Kunming, 650500, PR China}

\author{Shun-Cai Zhao\orcidA{}}
\email[Corresponding author: ]{zhaosc@kmust.edu.cn}
\affiliation{Center for Quantum Materials and Computational Condensed Matter Physics, Faculty of Science, Kunming University of Science and Technology, Kunming, 650500, PR China}
\affiliation{Department of Physics, Faculty of Science, Kunming University of Science and Technology, Kunming, 650500, PR China}

\author{Ya-Fang Tian}
\affiliation{Center for Quantum Materials and Computational Condensed Matter Physics, Faculty of Science, Kunming University of Science and Technology, Kunming, 650500, PR China}
\affiliation{Department of Physics, Faculty of Science, Kunming University of Science and Technology, Kunming, 650500, PR China}

\begin{abstract}
For efficient photovoltaic conversion, it is important to understand how quantum entropy-related quantities evolve during the photovoltaic process. In this study, using a double quantum dot (DQD) photocell model, we explored the dynamic quantum entropy-related parameters during the photovoltaic output. The findings demonstrate that the dynamic photovoltaic performance is compatible with quantum entropy-related parameters with varying tunneling coupling strengths, but at varied ambient temperatures, an opposing relationship is discovered between them. Hence, some thermodynamic criteria may be used to evaluate the photovoltaic process in this proposed photocell model. This work's merits include expanding our understanding of photoelectric conversion from a thermodynamic perspective as well as perhaps suggesting a new thermodynamic approach to efficient photoelectric conversion for DQD photocells.
\begin{description}
\item[Keywords]{Quantum entropy evolution; photovoltaic process; quantum dot photocell}
\end{description}
\end{abstract}

\maketitle
\section{Introduction}

Quantum photocells have been proposed as an excellent theoretical model\cite{scully2010quantum} for solar cells, and many schemes\cite{2011Enhancing,dorfman2011increasing,creatore2013efficient,zhao2019enhanced} for the efficient photovoltaic efficiency were carried out on the photocell. The efficiency losses arises from the radiative recombination\cite{shockley1961detailed}in the photoelectric conversion process owing to upward and downward transitions coexist simultaneously, are called intrinsic losses. And some researchers\cite{2011Enhancing,scully2011quantum,chen2020radiative} proposed some strategies to reduce radiative recombination, such as dipole-dipole interaction within molecules\cite{creatore2013efficient}, environmental noise-induced coherence\cite{scully2011quantum} and Fano interference effect\cite{dorfman2013photosynthetic}. In addition, there are some other factors that cause the efficiency losses in photovoltaic devices, such as series resistance\cite{kar2002series}, parasitic recombination and contact shadowing\cite{2013Recent}, they are theoretically avoidable and consequently are called extrinsic losses.

Quantum photocells just like classical heat engines, convert photon energy from the solar into electric energy. In order to clarify the physical correlation between photovoltaic process and thermodynamics performance, several studies\cite{scully2011quantum,li2021influence,dong2021thermodynamic} have attempted to reveal the photoelectric conversion process from the prospective of the thermodynamics in the quantum photocells. In the quantum thermodynamics approach, the photocell is usually described by a two-level atom and the generalized quantum master equations are used with the Jaynes-Cummings Hamiltonian\cite{gelbwaserklimovsky2017on,alicki2016solar}.
The interaction of a quantum system with its environment will generate correlations due to no perfect isolation between quantum system and its environment, which will cause loss of information\cite{bekenstein1973black}. Thus, the entropy-related quantities will quantifies the flow of information between the system and its environment. Not only that, but many natural phenomena described physical laws are based on entropy, such as the second law of thermodynamics\cite{attard2012nonequilibrium}, entropic uncertainty relations\cite{maassen1988generalized} and area laws in black holes\cite{lucabombelli1986quantum}.

However, few studies have sought to elucidate the efficient photoelectric conversion law from the thermodynamic perspective, particularly in light of the evolutions of entropy-related thermodynamic quantities. Therefore, for the sake of the new law of high efficient photoelectric conversion from the evolution of entropy-dependent quantities, we discuss some entropy-dependent thermodynamic quantities in the photovoltaic process of a proposed DQD photocell, and suggested some new strategies for the efficient photovoltaic efficiency in the DQD photocells.

The overall layout of this paper is as follows. In Section \ref{section2}, we briefly review the evolutions of some entropy-related thermodynamic quantities in an open quantum systems. In Section \ref{section3}, a DQD photocell model based on the master equation method is proposed, and we calculate its entropy-related thermodynamic quantities in the photovoltaic progress. Numerical simulation and analysis will be presented in section \ref{section4}, and we compare the photovoltaic characteristics and entropy evolution process under the same parameter conditions. Lastly, the summarization is given in Section \ref{section5}.

\section{Entropy evolution of an open quantum system}\label{section2}

This part starts some quantum entropy-related quantities of an open quantum system that exchanges entropy and heat with its environment. These thermodynamic quantities presented here will be applied to the proposed DQD photocell model. The second law of thermodynamics describes the irreversibility of kinetics, in which entropy is a fundamental quantity that is of wide interest in physics and information theory\cite{shannon1948a}. The von-Neumann entropy {S}(t) of the system in the state $\rho(t)$ is defined as

\begin{equation}
{S}(t)=-Tr[{\rho(t)}ln\rho(t)].\label{1}
\end{equation}

\noindent The entropy concept in thermodynamics S is written as S=${k}_{B}{S}(t)$. The most general formula for the second law of thermodynamics is Clausius' equations\cite{spohn1978entropy,bulnes2016quantum}, which indicated that the total entropy variation must be non-decreasing for the class of unital physical processes. Therefore, the total entropy change of the system and the environment (reservoir) must meet the relationship $\Delta{S}_{T}=\Delta{S}_{s}-\Delta{S}_{r} \ge 0$. The variation in entropy of the system, $\dot S_{s}$ over time is written as,

\begin{equation}
\dot{S}_{s}=-Tr[\dot\rho(t)ln\rho(t)+\rho(t)ln\dot\rho(t)]=-Tr[\dot\rho(t)ln\rho(t)],\label{2}
\end{equation}

\noindent in Eq.\ref{2} the normalization condition for population, $Tr[\dot\rho(t)=0]$ is used. The reversible entropy current from the environment to the system is given by\cite{alicki1979the},

\begin{equation}
\dot{S}_{r}=\beta Tr[\dot\rho(t)H_s(t)]=\beta\dot{Q}(t),\label{3}
\end{equation}

\noindent where $\beta$ = $\frac{1}{k_{B}T_{a}}$, and the heat current from the environment into the system $\dot{Q}(t)$ is equal to $Tr[\dot\rho(t)H_s(t)]$\cite{quan2007quantum,landsberg1980thermodynamic,boukobza2006thermodynamics}. Thus, the net entropy production rate $\sigma(t)$ of the system coupled to environments is written as follows\cite{spohn1978entropy},

\begin{equation}
\sigma(t)=\dot{S}_{s}-\beta\dot{Q}(t),  \label{4}
\end{equation}

\noindent Spohn demonstrated this to be an entirely positive map, such as for the Lindblad super operator\cite{spohn1978irreversible,alicki1979quantum,das2018fundamental}

\begin{equation}
\sigma(t)\ge 0,  \label{5}
\end{equation}

\noindent Eq.(\ref{5}) is Clausius¡¯s general formulation of the second law in differential form, which is valid at all times. Eq.(\ref{5}) may be written as,

\begin{equation}
\sigma\equiv-\frac{d}{dt}S(\rho(t)||\rho_{ss})\ge 0, \label{6}
\end{equation}

\noindent where $S(\rho(t)||\rho_{ss})\equiv Tr[\rho(t)(ln\rho(t)-ln\rho_{ss})]$ is the relative entropy of $\rho(t)$ with respect to the stationary state $\rho_{ss}$, which is called the second law of non-equilibrium quantum thermodynamics in the weak coupling limit.

\section{DQD photocell model in the photovoltaic process} \label{section3}

\begin{figure}[htp]\center
\includegraphics[scale=.85]{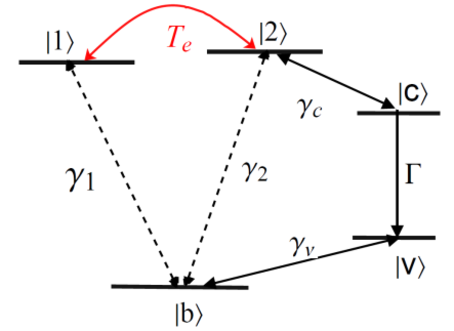}\hspace{0in}%
\caption{(Color online) Schematic a five-level DQD photocell model. Solar radiation continuously excites the electron transition $|b\rangle\leftrightarrow|1\rangle$($|2\rangle$) at the rate of $\gamma_{1}$($\gamma_{2}$). Ambient phonons mediate the low-energy transitions $|2\rangle\leftrightarrow|c\rangle$ and $|v\rangle\leftrightarrow|b\rangle$ at rates $\gamma_{c}$ and $\gamma_{v}$, respectively. States $|c\rangle$ and $|v\rangle$ are connected with an external terminal at the rate $\Gamma$ standing for the external load or electrical resistance. ${T}_{e}$ (Red double-arrow arc) depicts the tunneling coupling strength.}\label{Fig.1}
\end{figure}

For a DQD photocell model, let us now consider two vertically aligned QDs with different sizes, which was depicted as a five-level quantum system and composed of a donor and an acceptor in Fig.\ref{Fig.1}. Initially the DQD photocell was uncharged in the ground state $|b\rangle$, and the excited electrons and holes could tunnel between the two QDs\cite{2006Engineering} by tunneling ${T}_{e}$\cite{2022Enhancedaa}, after optical excitation from the transitions $|b\rangle\leftrightarrow|1\rangle$($|2\rangle$) which acts as the donor in this photocell. The energy levels marked with $|1\rangle$, $|2\rangle$ represent the electronic (excited) states of the photocell in Fig.\ref{Fig.1}, and $|b\rangle$ is the hole (ground) state. This photovoltaic conversion cycle begins with the absorption of solar photons with the average photon occupation numbers being ${n}_{1}$=$[exp(\frac{({E}_{1}-{E}_{b})}{{k}_{B}{T}_{h}})-1]^{-1}$ and ${n}_{2}$=$[exp(\frac{({E}_{2}-{E}_{b})}{{k}_{B}{T}_{h}})-1]^{-1}$ in the excited states $|1\rangle$ and $|2\rangle$ , respectively. ${E}_{i}$(i=b,1,2) is the intrinsic energy of the corresponding energy level and ${T}_{h}$ is the solar temperature. The electron-hole recombination rate in the transition $|b\rangle$$\leftrightarrow$$|1\rangle$($|2\rangle$) is represented by $\gamma_{1}$($\gamma_{2}$).

The second process involves the interaction of the DQD photocell with its surrounding environment (the reservoir) through the transformations $|2\rangle$$\leftrightarrow$$|c\rangle$ and $|v\rangle$$\leftrightarrow$$|b\rangle$.
The transition $|2\rangle$$\leftrightarrow$$|c\rangle$ with average phonon occupation numbers ${n}_{c}$=$[exp(\frac{({E}_{2}-{E}_{c})}{{k}_{B}{T}_{a}})-1]^{-1}$ transfers the excited electrons to the acceptor state, i.e., the charge separated state $|c\rangle$ at ambient environment temperature ${T}_{a}$\cite{dorfman2013photosynthetic}. The acceptor state $|c\rangle$ and the positively charged state $|v\rangle$ can be coupled to an external load. The excited electron is then thought to work through the transition $|c\rangle$$\rightarrow$$|v\rangle$ at a rate of $\Gamma$. Finally, the cycle is completed, causing the $|v\rangle$  state to decay back to the neutral ground state $|b\rangle$ at the rate $\gamma_{v}$ and with an average occupancy of ${n}_{v}$=$[exp(\frac{({E}_{v}-{E}_{b})}{{k}_{B}{T}_{a}})-1]^{-1}$  on the state $|v\rangle$.

Thus, as stated previously, a five-level system describing an opto-electrical conversion process is created. As shown in the following formula, the output voltage\cite{scully2010quantum,2011Enhancing,dorfman2011increasing}at the external load:

\begin{equation}
{eV}={E}_{c}-{E}_{v}+{k}_{B}{T}_{a}ln(\frac{\rho_{cc}}{\rho_{vv}}),\label{7}
\end{equation}

\noindent where, ${k}_{B}$ is the Boltzmann constant and $e$ is the fundamental charge at an ambient environment temperature ${T}_{a}$. $\rho_{cc}$ and $\rho_{vv}$ represent the population rates on the states $|c\rangle$ and $|v\rangle$, respectively. The output voltage $V$ over the external load is defined from the chemical energy difference\cite{creatore2013efficient} between the states $|c\rangle$ and $|v\rangle$. Then, the electric current\cite{zhao2019enhanced} $j$ through the external terminal is interpreted as,

\begin{equation}
{j}=e\Gamma\rho_{cc},           \label{8}
\end{equation}

\noindent $\Gamma$ is the transition rate $|c\rangle$$\rightarrow$$|v\rangle$. Therefore, the output power $P$ of DQD photocell can be written as,
\begin{equation}
{P}=jV.                         \label{9}
\end{equation}

\noindent Therefore, the photo-electric conversion efficiency of the DQD photocell system can be calculated as follows,

\begin{equation}
{\eta}=\frac{P}{P_{in}},\label{10}
\end{equation}

\noindent where the incident solar radiation power ${P}_{in}$ equals to j$\frac{({E}_{1}-{E}_{b})}{e}$.
From the perspective of quantum mechanics, the Hamiltonian of the total DQDs photocell system can be interpreted by four parts,

\begin{equation}
\hat{H}_{T}=\hat{H}_{0}+\hat{H}_{B}+\hat{V}_{hot}+\hat{V}_{cold}.\label{11}
\end{equation}

\noindent Where $\hat{H}_{0}$ is represented as the free Hamiltonian of the five-level system by the following form,

\begin{equation}
\hat{H}_{0}=\sum_{i=b,1,2,c,v}\hbar\omega_{i}|i\rangle\langle i|+{T}_{e}(|1\rangle\langle 2|+|2\rangle\langle 1|),\label{12}
\end{equation}

\noindent where $\omega_{i}$ is the transition frequency between different energy levels, and ${T}_{e}$ is the tunneling rate describing the tunneling effect between two quantum dots. For the sake of simplicity, we set $\hbar$ = ${1}$ in the following numerical calculations. The second term  in Eq.(\ref{11}), $\hat{H}_{B}$, depicts the free Hamiltonian of the environment reservoirs,

\begin{align}
\hat{H}_{B}=\sum_{k}\hbar\omega_{k}\hat{a}^{\dag}_{k}\hat{a}_{k}+\sum_{l}\hbar\omega_{l}\hat{b}^{\dag}_{l}\hat{b}_{l}+\sum_{m}\hbar\omega_{m}\hat{c}^{\dag}_{m}\hat{c}_{m},\label{13}
\end{align}

\noindent where $\hat{a}^{\dag}_{k}$($\hat{a}_{k}$) is the creation(annihilation) operator of thermal photons with the k-th frequency being $\omega_{k}$. The creation (annihilation) operators $\hat{b}^{\dag}_{l}$($\hat{b}_{l}$) and $\hat{c}^{\dag}_{m}$($\hat{c}_{m}$) of the ambient phononic environment with frequencies $\omega_{l}$ and $\omega_{m}$, respectively. Considering the weak coupling, the last two terms in Eq.(\ref{11}) describe the interactions between the ambient environment and the conduction band state($|1\rangle (|2\rangle)$) and valence band state($|b\rangle$) with their coupling coefficients being ${g}_{k}$, ${g}_{l}$, ${g}_{m}$,

\begin{align}
\hat{V}_{hot}=&\sum_{k}\hbar{g}_{k}(|1\rangle\langle b|+|2\rangle\langle b|)\hat{a}_{k}+H.c.,\label{14}\\
\hat{V}_{cold}=&\sum_{l}\hbar{g}_{l}|2\rangle\langle c|\hat{b}_{l}+\sum_{m}\hbar{g}_{m}|v\rangle\langle b|\hat{c}_{m}+H.c.\label{15}
\end{align}

Under the second-order perturbation treatment, using the Born-Markov and Weisskopf-Wigner approximations\cite{wang1974wigner,wertnik2018optimizing}, the dynamics of the five-level DQD system can be described by the Lindblad master equation\cite{lindblad1975completely,lindblad1976generators},
\begin{equation}
\frac{d\hat\rho}{dt}=-i[\hat{H}_{0},\hat\rho]+\mathscr{L}_{H}\hat{\rho}+\mathscr{L}_{c}\hat{\rho}+\mathscr{L}_{v}\hat{\rho}+\mathscr{L}_{\Gamma}\hat{\rho}.\label{16}
\end{equation}

\noindent Where $\hat{\rho}$ is the reduced density operators, and the expression of $\mathscr{L}_{i}\hat{\rho}(i\!=\!H,c,v,\Gamma)$ is listed as follows,

\begin{align}
\mathscr{L}_{H}\hat{\rho}&=\sum_{i=1,2}\frac{\gamma_{i}}{2}[(n_{i}+1)(2\hat{\sigma}_{bi}\hat{\rho}\hat{\sigma}_{bi}^{\dag}-\hat{\sigma}_{bi}^{\dag}\hat{\sigma}_{bi}\hat{\rho}-\hat{\rho}\hat{\sigma}_{bi}^{\dag}\hat{\sigma}_{bi})
                                                +n_{i}(2\hat{\sigma}_{bi}^{\dag}\hat{\rho}\hat{\sigma}_{bi}-\hat{\sigma}_{bi}\hat{\sigma}_{bi}^{\dag}\hat{\rho}-\hat{\rho}\hat{\sigma}_{bi}\hat{\sigma}_{bi}^{\dag})],\label{17}\\
\mathscr{L}_{c}\hat{\rho}&=\frac{\gamma_{c}}{2}[(n_{c}+1)(2\hat{\sigma}_{c2}\hat{\rho}\hat{\sigma}_{c2}^{\dag}-\hat{\sigma}_{c2}^{\dag}\hat{\sigma}_{c2}\hat{\rho}-\hat{\rho}\hat{\sigma}_{c2}^{\dag}\hat{\sigma}_{c2})
                                                +n_{c}(2\hat{\sigma}_{c2}^{\dag}\hat{\rho}\hat{\sigma}_{c2}-\hat{\sigma}_{c2}\hat{\sigma}_{c2}^{\dag}\hat{\rho}-\hat{\rho}\hat{\sigma}_{c2}\hat{\sigma}_{c2}^{\dag})],\label{18}\\
\mathscr{L}_{v}\hat{\rho}&=\frac{\gamma_{v}}{2}[(n_{v}+1)(2\hat{\sigma}_{bv}\hat{\rho}\hat{\sigma}_{bv}^{\dag}-\hat{\sigma}_{bv}^{\dag}\hat{\sigma}_{bv}\hat{\rho}-\hat{\rho}\hat{\sigma}_{bv}^{\dag}\hat{\sigma}_{bv})+n_{v}(2\hat{\sigma}_{bv}^{\dag}\hat{\rho}\hat{\sigma}_{bv}-\hat{\sigma}_{bv}\hat{\sigma}_{bv}^{\dag}\hat{\rho}-\hat{\rho}\hat{\sigma}_{bv}\hat{\sigma}_{bv}^{\dag})],\label{19}\\
\mathscr{L}_{\Gamma}\hat{\rho}&=\frac{\Gamma}{2}(2\hat{\sigma}_{vc}\hat{\rho}\hat{\sigma}_{vc}^{\dag}-\hat{\sigma}_{vc}^{\dag}\hat{\sigma}_{vc}\hat{\rho}-\hat{\rho}\hat{\sigma}_{vc}^{\dag}\hat{\sigma}_{vc})\label{20}
\end{align}

\noindent where $\mathscr{L}_{H}\hat{\rho}(i\!=\!1,2)$ labels the transition $|b\rangle$$\leftrightarrow$$|i\rangle$ with Pauli operator $\hat{\sigma}_{bi}$=$|b\rangle\langle i|$, spontaneous decay rates $\gamma_{i}$ and average photon number ${n}_{i}$. $\mathscr{L}_{c}\hat{\rho}$ represents the transition $|2\rangle$$\leftrightarrow$$|c\rangle$ with Pauli operator $\hat{\sigma}_{c2}$=$|c\rangle\langle 2|$, spontaneous decay rate $\gamma_{c}$ and corresponding phonon occupation number ${n}_{c}$. Similarly, $\mathscr{L}_{v}\hat{\rho}$ denotes another interaction between the DQD photocell system and the ambient reservoir through the transition $|v\rangle$$\leftrightarrow$$|b\rangle$ with Pauli operator $\hat{\sigma}_{bv}$=$|b\rangle\langle v|$, decay rate $\gamma_{v}$ and average phonon number ${n}_{v}$. Eventually, $\mathscr{L}_{\Gamma}\hat{\rho}$ describes the transition $|c\rangle$$\rightarrow$$|v\rangle$ by the rate $\Gamma$, and Pauli operator $\hat{\sigma}_{vc}$=$|v\rangle\langle c|$.

\par In consequence, in the Schr{\"o}dinger picture, the expressions of the complete population evolution are expressed according to Eq.(\ref{16}) $\sim$ Eq.(\ref{20}) as follows,

\begin{align}
\dot{\rho}_{11}=&-i{T}_{e}(\rho_{21}-\rho_{12})-\gamma_{1}[(n_{1}+1)\rho_{11}-n_{1}\rho_{bb}],\label{21}\\
\dot{\rho}_{22}=&i{T}_{e}(\rho_{21}-\rho_{12})-\gamma_{2}[(n_{2}+1)\rho_{22}-n_{2}\rho_{bb}]-\gamma_{c}[(n_{c}+1)\rho_{22}-n_{c}\rho_{cc}],\label{22}\\
\dot{\rho}_{12}=&-i\hbar(\omega_{1}-\omega_{2})\rho_{12}-i{T}_{e}(\rho_{22}-\rho_{11})-\frac{\rho_{12}}{2}[\gamma_{1}(n_{1}+1)+\gamma_{2}(n_{2}+1)+\gamma_{c}(n_{c}+1)],\label{23}\\
\dot{\rho}_{21}=&-i\hbar(\omega_{2}-\omega_{1})\rho_{21}-i{T}_{e}(\rho_{11}-\rho_{22})-\frac{\rho_{21}}{2}[\gamma_{1}(n_{1}+1)+\gamma_{2}(n_{2}+1)+\gamma_{c}(n_{c}+1)], \label{24}\\
\dot{\rho}_{cc}=&\gamma_{c}[(n_{c}+1)\rho_{22}-n_{c}\rho_{cc}]-\Gamma\rho_{cc},\label{25}\\
\dot{\rho}_{vv}=&\Gamma\rho_{cc}-\gamma_{v}[(n_{v}+1)\rho_{vv}-n_{v}\rho_{bb}]   \label{26}
\end{align}

\noindent where $\rho_{ii}$ represents the diagonal element, and $\rho_{ij}$ is the non-diagonal element of the corresponding state. Therefore, according to Eq.(\ref{1})$\sim$ Eq.(\ref{6}), the entropy $S_{A}(t)$ and the entropy change of the acceptor $\dot S_{A}(t)$ over time can be calculated in the DQD photocell. Simultaneously, with the steady-state solution $\rho^{s}_{ii}$, the total entropy $S(t)$, heat current $\dot{Q}(t)$ and the net entropy production rate $\sigma(t)$ of the photovoltaic system can be deduced as follows, respectively.

\begin{align}
{S}_{A}(t)=&-(\rho_{cc}ln\rho_{cc}+\rho_{vv}ln\rho_{vv}),  \label{28}\\
\dot{S}_{A}(t)=&-[(\gamma_{c}[(n_{c}+1)\rho_{22}-n_{c}\rho_{cc}]-\Gamma\rho_{cc})(ln\rho_{cc}+{1})+
              (\Gamma\rho_{cc}-\gamma_{v}[(n_{v}+1)\rho_{vv}\nonumber\\
              &-n_{v}\rho_{bb}])(ln\rho_{vv}+{1})], \label{29}\\
{S}(t)=&-[\rho_{11}ln\rho_{11}+\rho_{22}ln\rho_{22}+\rho_{bb}ln\rho_{bb}+\rho_{cc}ln\rho_{cc}+\rho_{vv}ln\rho_{vv}]\\
\dot{Q}(t)=& \hbar \omega_{1} \dot{\rho}_{11}+ \hbar \omega_{2} \dot{\rho}_{22}+ \hbar \omega_{b} \dot{\rho}_{bb}+ \hbar \omega_{c} \dot{\rho}_{cc}+ \hbar \omega_{v} \dot{\rho}_{vv}+ {T}_{e}(\dot{\rho}_{21}+\dot{\rho}_{12}), \label{30}\\
\sigma(t)=&-[-i{T}_{e}(\rho_{21}-\rho_{12})-\gamma_{1}[(n_{1}+1)\rho_{11}-n_{1}\rho_{bb}]](ln\rho_{11}-
          ln\rho^{s}_{11})-[i{T}_{e}(\rho_{21}-\rho_{12})\nonumber{}\\
         &-\gamma_{2}[(n_{2}+1)\rho_{22}-n_{2}\rho_{bb}]-\gamma_{c}[(n_{c}+1)\rho_{22}-n_{c}\rho_{cc}]](ln\rho_{22}-ln\rho^{s}_{22})-[\gamma_{c}[(n_{c}+1)\rho_{22}-n_{c}\rho_{cc}]\nonumber{}\\
         &-\Gamma\rho_{cc}](ln\rho_{cc}-ln\rho^{s}_{cc})-[\Gamma\rho_{cc}-\gamma_{v}[(n_{v}+1)\rho_{vv}-n_{v}\rho_{bb}]](ln\rho_{vv}-ln\rho^{s}_{vv})-[\gamma_{1}[(n_{1}+1)\rho_{11}\nonumber\\
         &-n_{1}\rho_{bb}]+\gamma_{2}[(n_{2}+1)\rho_{22}-n_{2}\rho_{bb}]+\gamma_{v}[(n_{v}+1)\rho_{vv}-n_{v}\rho_{bb}]](ln\rho_{bb}-ln\rho^{s}_{bb}).\label{31}
\end{align}

\section{Numerical simulation and analysis}
\label{section4}
\begin{table}
\begin{center}
\caption{Typical parameters used in this work.}
\label{Table 1}
\vskip 0.2cm\setlength{\tabcolsep}{0.5cm}
\begin{tabular}{ccc}
\hline
\hline
                       & Values                     &Units\\
\hline
\(\gamma_{1}\)         & 0.12 \(\gamma_{0}\)         & eV  \\
\(\gamma_{2}\)         & 0.1 \(\gamma_{0}\)          & eV  \\
\(\gamma_{c}\)         & 1.65 \(\gamma_{0}\)         & eV  \\
\(\gamma_{v}\)         & 0.48 \(\gamma_{0}\)         & eV  \\
\(\Gamma\)             & 1.25\(\gamma_{0}\)          & eV  \\
$T_{h}$                & 0.5                        & eV  \\
\(\omega_{b}\)         & 0                          & eV  \\
$E_{1}-E_{c}$          & 1.28                       & eV  \\
$E_{2}-E_{c}$         & 0.02                       & eV  \\
$E_{c}-E_{v}$          & 0.50                       & eV  \\
$E_{v}-E_{0}$          & 0.20                       & eV  \\
\(\gamma_{0}\)        & \(10^{-3}\)                & scale unit \\
\hline
\hline
\end{tabular}
\end{center}
\end{table}

The evolution behavior of quantum entropy-related parameters throughout the photoelectric conversion process may indicate the properties of photoelectric conversion if we consider the DQD photocell to be a thermodynamic system. As seen from the perspective of quantum thermodynamic entropy, this may give humans the opportunity to maximize the photoelectric conversion efficiency. Therefore, in this paper, we will investigate the evolution of quantum thermodynamic quantities associated with the entropy in the photoelectric conversion process of DQD photocell and compare them to the photovoltaic physical quantities, so as to reveal the equivalence of entropy evolution and charge transfer in the photovoltaic process.

In this proposed DQD photocell model, the cyclic operation of the donor-acceptor DQD photocell can be performed by the initial absorbing photons in the donor part, and the electrons become excited with the transitions from the valence band (VB) state $|b\rangle$ to the conduction band (CB) states $|1\rangle$ ($|2\rangle$). Considering the average absorbed photons mentioned by some previous work\cite{oh2019efficiency,tomasi2020classification}, the number of photons absorbed per unit time are set as $n_{1}$($n_{2}$)=0.01, which corresponds to about 2.3 $eV$ band gap with a black body at 5800K. The phonon vibration makes the excited electrons at the donor transfer to the acceptor state $|c\rangle$. The acceptor is coupled to an external load and the electric current represented by the transition decays from the state $|c\rangle$ to the state $|v\rangle$. All the other operating parameters referred to some experimental work\cite{2006Engineering,2005Experimental,2014Noise} are listed in the Table \ref{Table 1}.

\subsection{ Entropy dependent the tunneling coupling strength Te}

\begin{figure}[htp]
\center
\includegraphics[scale=.85]{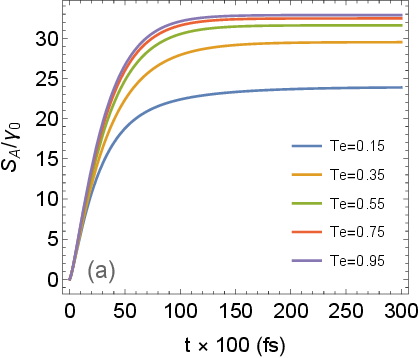}\hspace{0in}%
\includegraphics[scale=.85]{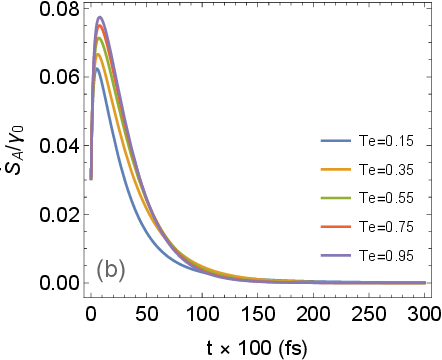}\hspace{0in}%
\includegraphics[scale=.85]{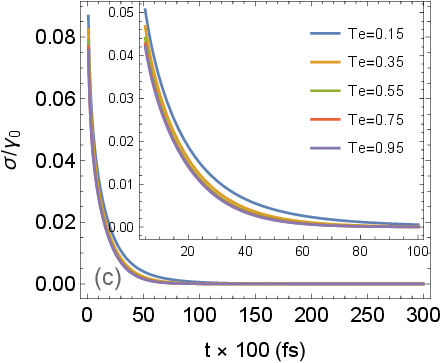}\hspace{0in}%
\includegraphics[scale=.85]{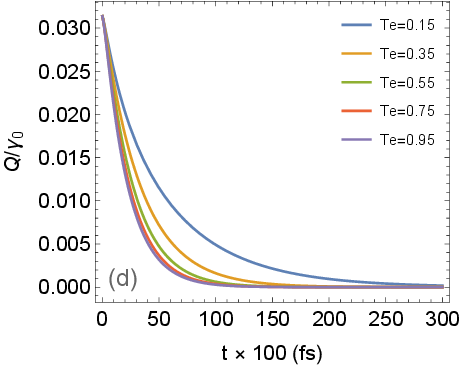}
\caption{(Color online) Dynamic evolution of quantities associated with the entropy dependent different tunneling coupling strength $T_{e}$ in this proposed DQD photocell. (a) The entropy $S_{A}$ of the acceptor part, (b) the entropy flow $\dot{S}_{A}$ from the donor to acceptor, (c) the net entropy production rate $\sigma$, (d) the heat current $\dot{Q}$ from the environment into the DQD photocell system with other parameters taken from Table \ref{Table 1}.}\label{Fig.2}
\end{figure}

\par The tunnel coupling coefficient $T_{e}$ between two quantum dots is an essential parameter that influences photovoltaic performance in the DQD photocell model. So, next, the dynamical properties of entropy-dependent quantum thermodynamic quantities will be discussed by the tunnel coupling coefficient $T_{e}$ at an ambient temperature $T_{a}$=0.026eV. As for a thermodynamic system, the second law of thermodynamics tells us that its entropy is always non-negative. In Fig.\ref{Fig.2}(a), taking the curve with $T_{e}$=0.15 as an example, the entropy $S_A$ of the acceptor steadily grows over time with the largest increment in the period of [0, 100$\times$100(fs)], and gradually achieves a steady value. Furthermore, we notice that the increment in stable entropy decreases as $T_e$ increases by 0.2 from 0.15 to 0.95 in Fig.\ref{Fig.2}(a). The steady entropy $S_A$ signifies that the DQD system has reached thermal equilibrium with its surroundings, which may denote that a steady photovoltaic output is obtained. The change in the entropy of the acceptor over time $\dot{S}_A$ implies the entropy flowing from the donor to acceptor, which further demonstrates the thermal equilibrium with its surroundings as  $t\ge 100\times$100(fs) in Fig.\ref{Fig.2}(b).The thermal equilibrium is also demonstrated by the asymptotic to zero net entropy production rate $\sigma$ in Fig.\ref{Fig.2}(c) and by the zero heat current $\dot{Q}$ from the environment into the DQD photocell system in Fig.\ref{Fig.2}(d).

\begin{figure}[htp]
\center
\includegraphics[scale=.85]{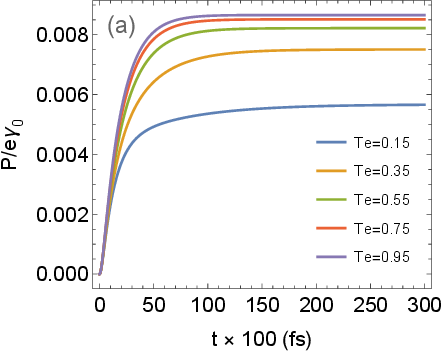 }\hspace{0in}%
\includegraphics[scale=.85]{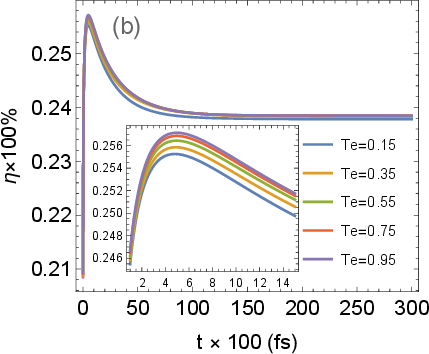 }\hspace{0in}%
\caption{(Color online) Dynamic evolution of the photovoltaic quantities dependent different tunneling coupling strength $T_{e}$ in this proposed DQD photocell. (a) The output power $P$, (b) the photoelectric conversion efficiency $\eta$ of the DQD photocell system with other parameters taken the same to those in Fig.\ref{Fig.2}.}\label{Fig.3}
\end{figure}

As previously indicated, the steady-state photoelectric conversion in the photovoltaic process may be predicted by the thermal equilibrium between the DQD photocell system and the surrounding environment. Fig.\ref{Fig.3} shows the dynamic evolution of the output power $P$ and photoelectric conversion efficiency $\eta$ in the photovoltaic process when the tunneling coupling strength $T_{e}$ is engineered. The curves in Fig.\ref{Fig.3}(a) show a trend similar to that in Fig.\ref{Fig.2}(a). The output power $P$ increase gradually over time, peaking at [0, 100$\times$100(fs)], before exhibiting a tendency toward stability. The output powers steadily increase but their amplitudes gradually decrease with the increment of the tunnel coupling coefficient $T_{e}$. The horizontal power output curves clearly illustrate these in Fig.\ref{Fig.3}(a).

The photoelectric conversion efficiency $\eta$ is a crucial metric to assess a photocell photovoltaic performance. The dynamic evolution of $\eta$ is plotted by the curves for different tunneling coupling ${T}_{e}$  in Fig.\ref{Fig.3}(b). As shown by the curves of insert illustration in Fig.\ref{Fig.3}(b), ${T}_{e}$ plays a positive role in the photoelectric conversion efficiency $\eta$, i.e., $\eta$ increases with the increase of ${T}_{e}$.
What's more, we notice the curves in Fig.\ref{Fig.3}(b) for the efficiency $\eta$ are consistent with those in Fig.\ref{Fig.2}(b). As a result of examining the quantum thermodynamic quantities and photovoltaic characteristic parameters in this DQD photocell model, we have come to the conclusion of thermodynamic criteria for the photovoltaic process when the DQD photocell model is regulated by the tunnel coupling coefficient $T_{e}$.

\subsection{ Entropy dependent the ambient temperature Ta}

\begin{figure}[htp]
\center
\includegraphics[scale=.85]{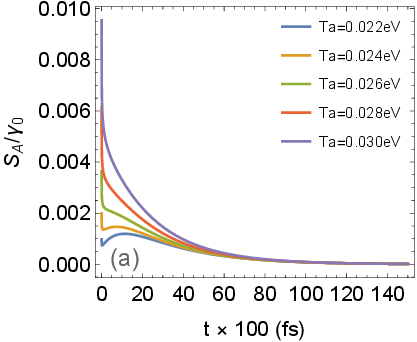 }\hspace{0in}%
\includegraphics[scale=.85]{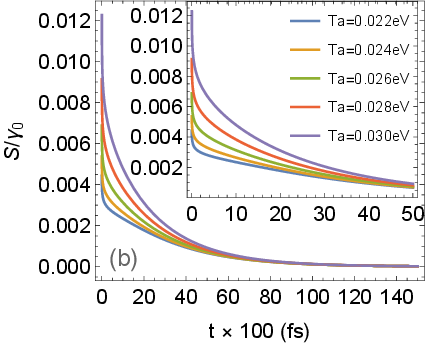 }\hspace{0in}%
\includegraphics[scale=.85]{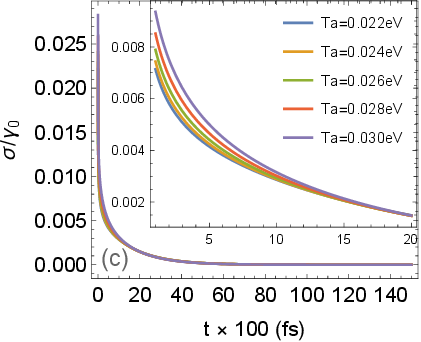 }\hspace{0in}%
\includegraphics[scale=.85]{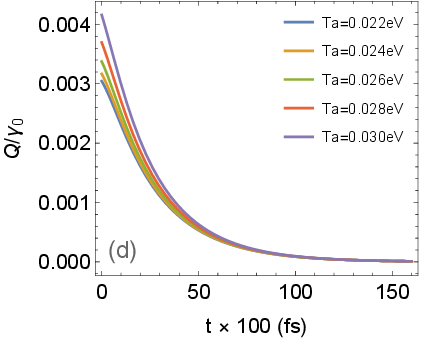}\hspace{0in}
\caption{(Color online) Dynamic evolution of quantities associated with the entropy versus different ambient temperatures $T_{a}$ in this proposed DQD photocell. (a) The entropy flow $\dot{S}_{A}$ from the donor to acceptor, (b) the entropy flow $\dot{S}$ of the DQD photocell system, (c) the net entropy production rate $\sigma$, (d) the heat current $\dot{Q}$ from the environment into the DQD photocell system with $\gamma_{1}$=0.012$\gamma_{0}$, $\gamma_{2}$=0.01$\gamma_{0}$, $E_{1}-E_{c}$=0.98eV, $E_{2}-E_{c}$=0.22eV, other parameters are the same to those in Fig.\ref{Fig.2}.}\label{Fig.4}
\end{figure}

Ambient temperature is one of the key factors affecting the performance of photovoltaic devices, however it is unclear how the temperature affects the quantum thermodynamic quantities of the devices. The following will examine the evolution behavior of quantum thermodynamic quantities as well as the features of the photovoltaic parameters at various ambient temperatures $T_a$ in this DQD photocell. In the following discussion, the tunneling coupling strength $T_e$ is set as 0.55. It should be noticed that the time evolution curves for quantum thermodynamic parameters in Fig.\ref{Fig.4}, the partial entropy change of the acceptor over time $\dot{S}_A$, the total entropy change of the system over time $\dot{S}$, the heat changing over time $\dot{Q}$ and the net entropy production rate $\sigma$, all present horizontal curves and are close to zero in slope from (a),(b),(c) to (d). This indicates that the DQD photocell will eventually reach the thermodynamic equilibrium with the surrounding environment at different ambient temperatures. However, in the period of [0, 40$\times$100(fs)], especially in the inner illustrations in Fig.\ref{Fig.4}(a) and (b), it is obvious that the higher the external temperature $T_a$ is at different moments, the greater the quantum thermodynamic quantities will be. Even though the system eventually reaches thermodynamic equilibrium with the environment at different ambient temperatures.

\begin{figure}[htp]
\center
\includegraphics[scale=.85]{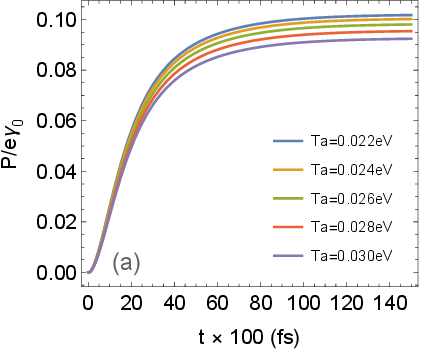 }\hspace{0in}%
\includegraphics[scale=.85]{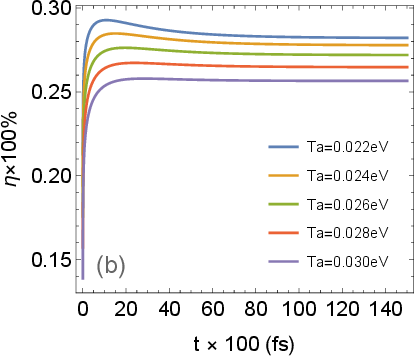 }\hspace{0in}%
\caption{(Color online)  Dynamic evolution of the photovoltaic quantities dependent different ambient temperatures $T_{a}$ in this proposed DQD photocell. (a) The output power $P$, (b) the photoelectric conversion efficiency $\eta$ of the DQD photocell system with other parameters taken the same to those in Fig.\ref{Fig.4}.} \label{Fig.5}
\end{figure}

Accordingly, the photovoltaic characteristic quantities are plotted in Fig.\ref{Fig.5} under the same parameter conditions to Fig.\ref{Fig.4}. Compared with the tunnel coupling coefficient $T_{e}$, $T_{a}$ plays a negative role in photovoltaic characteristic quantities. As can be seen from the curves in Fig.\ref{Fig.5}(a), the steady-state values of the output power $P$ decrease with the increments of $T_{a}$. Moreover, the speed of reaching the output peak power is less than that of $T_{e}$ in Fig.\ref{Fig.3}(a). The curves of photoelectric conversion efficiency $\eta$ in Fig.\ref{Fig.5}(b) demonstrates the identical conclusion due to the different ambient temperature $T_{a}$.

Before concluding this paper, we would like to make some comments. First of all, we believe that the physical nature of transforming photons to output electricity in the photovoltaic process is essentially similar to that of heat absorption to useful work in the heat engine, which results in the thermodynamic evolution process is related to the photovoltaic quantities. So that the photovoltaic quantities directly reproduces the thermodynamic features in the DQD photocell proposed by this work. Second, despite the fact that there are several approaches\cite{dong2021thermodynamic} to study the thermodynamic evolution of a DQD photocell and numerous quantum thermodynamic variables to assess its thermodynamic behaviors, this paper holds that these approaches are not the goal of this work. The purpose of this study is to determine the photovoltaic evolution's thermodynamic criteria using some specified quantum thermodynamic variables, and the attainment of the primary goal of this study is not affected by the  quantum thermodynamic values used here.

\section{Conclusions and remarks} \label{section5}
In conclusion, we chose two typical factors, tunneling coupling strength $T_e$ and ambient temperature $T_{a}$, which have an impact on the evolution behavior of quantum entropy-related parameters and the photovoltaic performance of DQD photocells. The results show that the evolution of entropy-related parameters is consistent with photovoltaic performance in this DQD photocell with varying $T_e$, whereas an inverse relationship is revealed between the evolution of entropy-related parameters and photovoltaic performance at varying ambient temperature $T_{a}$. In particular, when the ambient temperature is adjusted, the photovoltaic performance falls with the growth of $T_{a}$ while the entropy-related quantum thermodynamic quantities increase. The law discovered in this study could one day offer a thermodynamic plan for effective photoelectric conversion for DQD photocells.

\section{acknowledgments}

We offer our thanks for the financial support from the National Natural Science Foundation of China (Grant Nos. 62065009 and 61565008), and Yunnan Fundamental Research Projects, China (Grant No. 2016FB009).

\section*{Conflict of Interest}

The authors declare that they have no conflict of interest. This article does not contain any studies with human participants or animals performed by any of the authors. Informed consent was obtained from all individual participants included in the study.

\bibliography{reference}

\begin{thebibliography}{10}

\bibitem{scully2010quantum}
M.~O. Scully.
\newblock Quantum photocell: Using quantum coherence to reduce radiative
  recombination and increase efficiency.
\newblock {\em Phys. Rev. Lett.}, 104(20):207701, 2010.

\bibitem{2011Enhancing}
A.~A. Svidzinsky, K.~E. Dorfman, and M.~O. Scully.
\newblock Enhancing photovoltaic power by fano-induced coherence.
\newblock {\em Phys. Rev. A.}, 84(5):6140, 2011.

\bibitem{dorfman2011increasing}
K.~E. Dorfman, M.~B. Kim, and A.~A. Svidzinsky.
\newblock Increasing photocell power by quantum coherence induced by external
  source.
\newblock {\em Phys. Rev. A.}, 84(5):053829, 2011.

\bibitem{creatore2013efficient}
C.~Creatore, M.~A. Parker, S.~Emmott, and A.~W. Chin.
\newblock Efficient biologically inspired photocell enhanced by delocalized
  quantum states.
\newblock {\em Phys. Rev. Lett.}, 111(25):253601, 2013.

\bibitem{zhao2019enhanced}
S.~C. Zhao and J.~Y. Chen.
\newblock Enhanced quantum yields and efficiency in a quantum dot photocell
  modeled by a multi-level system.
\newblock {\em New J. Physics.}, 21(10):103015, 2019.

\bibitem{shockley1961detailed}
W.~Shockley and H.~J. Queisser.
\newblock Detailed balance limit of efficiency of p-n junction solar cells.
\newblock {\em J. Appl. Phys.}, 32(3):510, 1961.

\bibitem{scully2011quantum}
M.~O. Scully, K.~R. Chapin, K.~E. Dorfman, M.~B. Kim, and A.~Svidzinsky.
\newblock Quantum heat engine power can be increased by noise-induced
  coherence.
\newblock {\em PNAS}, 108(37):15097, 2011.

\bibitem{chen2020radiative}
J.~Y. Chen and S.~C. Zhao.
\newblock Radiative recombination rate suppressed in a quantum photocell with
  three electron donors.
\newblock {\em Europ. Phys. J. Plus.}, 135(1):92, 2020.

\bibitem{dorfman2013photosynthetic}
K.~E. Dorfman, D.~V. Voronine, S.~Mukamel, and M.~O. Scully.
\newblock Photosynthetic reaction center as a quantum heat engine.
\newblock {\em PNAS}, 110(8):2746, 2013.

\bibitem{kar2002series}
G.~S. Kar, S.~Maikap, S.~K. Banerjee, and S.~K. Ray.
\newblock Series resistance and mobility degradation factor in c-incorporated
  sige heterostructure p-type metal--oxide semiconductor field-effect
  transistors.
\newblock {\em Semiconduct. Sci. and Tech.}, 17(9):938, 2002.

\bibitem{2013Recent}
Y.~Kajiyama, K.~Joseph, K.~Kajiyama, S.~Kudo, and H.~Aziz.
\newblock Recent progress on the vacuum deposition of oleds with feature sizes
  $\leq$ 20 $\mu$m using a contact shadow mask patterned in-situ by laser
  ablation.
\newblock {\em Organic Light Emitting Materials and Devices XVII.},
  8829:101--106, 2013.

\bibitem{li2021influence}
L.~F. Li and S.~C. Zhao.
\newblock Influence of the coupled-dipoles on photosynthetic performance in a
  photosynthetic quantum heat engine.
\newblock {\em Chin. Phys. B}, 30(4):044215, 2021.

\bibitem{dong2021thermodynamic}
H.~Dong, A.~Ghosh, M.~O. Scully, and G.~Kurizki.
\newblock Thermodynamic bounds on work extraction from photocells and
  photosynthesis.
\newblock {\em Europ. Phys. J. Spec. Topics.}, 230:873, 2021.

\bibitem{gelbwaserklimovsky2017on}
D.~Gelbwaser-Klimovsky and A.~Aspuru-Guzik.
\newblock On thermodynamic inconsistencies in several photosynthetic and solar
  cell models and how to fix them.
\newblock {\em Chemical Science}, 8(2):1008, 2017.

\bibitem{alicki2016solar}
R.~Alicki, D.~Gelbwaser-Klimovsky, and K.~Szczygielski.
\newblock Solar cell as a self-oscillating heat engine.
\newblock {\em J. Phys. A}, 49(1):015002, 2016.

\bibitem{bekenstein1973black}
J.~D. Bekenstein.
\newblock Black holes and entropy.
\newblock {\em Phys. Rev. D.}, 7(8):2333, 1973.

\bibitem{attard2012nonequilibrium}
P.~Attard.
\newblock {\em Non-equilibrium thermodynamics and statistical mechanics :
  foundations and applications}.
\newblock Oxford Univiversity Press, 2012.

\bibitem{maassen1988generalized}
H.~Maassen and J.~B.~M. Uffink.
\newblock Generalized entropic uncertainty relations.
\newblock {\em Phys. Rev. Lett.}, 60(12):1103, 1988.

\bibitem{lucabombelli1986quantum}
L.~Bombelli, R.~K. Koul, J.~Lee, and R.~D. Sorkin.
\newblock Quantum source of entropy for black holes.
\newblock {\em Physical Review D.}, 34(2):373, 1986.

\bibitem{shannon1948a}
C.~E. Shannon.
\newblock A mathematical theory of communication.
\newblock {\em Bell system technical journal}, 27(3):379, 1948.

\bibitem{spohn1978entropy}
H.~Spohn.
\newblock Entropy production for quantum dynamical semigroups.
\newblock {\em J. Math. Phys.}, 19(5):1227, 1978.

\bibitem{bulnes2016quantum}
G.~Bulnes~Cuetara, M.~Esposito, and G.~Schaller.
\newblock Quantum thermodynamics with degenerate eigenstate coherences.
\newblock {\em Entropy.}, 18(12):447, 2016.

\bibitem{alicki1979the}
R.~Alicki.
\newblock The quantum open system as a model of the heat engine.
\newblock {\em Journal of Physics A: Mathematical and General.}, 12(5):L103,
  1979.

\bibitem{quan2007quantum}
H.~T. Quan, Y.~X. Liu, C.~P. Sun, and F.~Nori.
\newblock Quantum thermodynamic cycles and quantum heat engines.
\newblock {\em Phys. Rev. E.}, 76(3):031105, 2007.

\bibitem{landsberg1980thermodynamic}
P.~T. Landsberg and G.~Tonge.
\newblock Thermodynamic energy conversion efficiencies.
\newblock {\em J. App. Phys.}, 51(7):R1--R20, 1980.

\bibitem{boukobza2006thermodynamics}
E.~Boukobza and D.~J. Tannor.
\newblock Thermodynamics of bipartite systems: Application to light-matter
  interactions.
\newblock {\em Phys. Rev. A.}, 74(6):063823, 2006.

\bibitem{spohn1978irreversible}
H.~Spohn and J.~L. Lebowitz.
\newblock Irreversible thermodynamics for quantum systems weakly coupled to
  thermal reservoirs.
\newblock {\em Adv. in Chem. Phys.: For Ilya Prigogine}, page 109, 1978.

\bibitem{alicki1979quantum}
R.~Alicki.
\newblock The quantum open system as a model of the heat engine.
\newblock {\em J. Phys. A: Math. and Gen.}, 12(5):L103, 1979.

\bibitem{das2018fundamental}
S.~Das, S.~Khatri, G.~Siopsis, and M.~M. Wilde.
\newblock Fundamental limits on quantum dynamics based on entropy change.
\newblock {\em J. Math. Phys.}, 59(1):012205, 2018.

\bibitem{2006Engineering}
A.~S. Bracker, M.~Scheibner, M.~F. Doty, E.~A. Stinaff, I.~V. Ponomarev, J.~C.
  Kim, L.~J. Whitman, T.~L. Reinecke, and D.~Gammon.
\newblock Engineering electron and hole tunneling with asymmetric inas quantum
  dot molecules.
\newblock {\em Appl. Phys. Lett.}, 89(23):233110, 2006.

\bibitem{2022Enhancedaa}
J.~Lira, J.~M. Villas-Boas, L.~Sanz, and A.~M. Alcalde.
\newblock Enhanced solar photocurrent using a quantum-dot molecule.
\newblock {\em J. Opt. Society of America, B. Optical Physics.}, 39(8):2047,
  2022.

\bibitem{wang1974wigner}
Y.~K. Wang and I.~C. Khoo.
\newblock On the wigner-weisskopf approximation in quantum optics.
\newblock {\em Opt. Comm.}, 11(4):323, 1974.

\bibitem{wertnik2018optimizing}
M.~Wertnik, A.~Chin, F.~Nori, and N.~Lambert.
\newblock Optimizing co-operative multi-environment dynamics in a
  dark-state-enhanced photosynthetic heat engine.
\newblock {\em The Journal of chemical physics}, 149(8):084112, 2018.

\bibitem{lindblad1975completely}
G.~Lindblad.
\newblock Completely positive maps and entropy inequalities.
\newblock {\em Comm. in Math. Phys.}, 40:147, 1975.

\bibitem{lindblad1976generators}
G.~Lindblad.
\newblock On the generators of quantum dynamical semigroups.
\newblock {\em Comm. in Math. Phys.}, 48:119, 1976.

\bibitem{oh2019efficiency}
S.~Oh.
\newblock Efficiency and power enhancement of solar cells by dark states.
\newblock {\em Phys. Lett. A}, 383(28):125857, 2019.

\bibitem{tomasi2020classification}
S.~Tomasi and I.~Kassal.
\newblock Classification of coherent enhancements of light-harvesting
  processes.
\newblock {\em The journal of physical chemistry letters.}, 11(6):2348, 2020.

\bibitem{2005Experimental}
A.~Luque, A.~Mart¨ª, N.~L¨®pez, E.~Antol¨ªn, and J.~L. Balenzategui.
\newblock Experimental analysis of the quasi-fermi level split in quantum dot
  intermediate-band solar cells.
\newblock {\em Appl. Phys. Lett.}, 87(8):1246, 2005.

\bibitem{2014Noise}
E.~R. Bittner and C.~Silva.
\newblock Noise-induced quantum coherence drives photo-carrier generation
  dynamics at polymeric semiconductor heterojunctions.
\newblock {\em Nat. Comm.}, 5(1):3119, 2014.

\end{thebibliography}
\bibliographystyle{unsrt}
\end{document}